\pdfoutput=1
\documentclass[12pt]{article}

\usepackage[T1]{fontenc}
\usepackage[utf8]{inputenc}
\usepackage[english]{babel}

\usepackage{amsmath}
\usepackage{amssymb}

\usepackage[natbibapa,nodoi]{apacite}
\usepackage{authblk}

\usepackage{bbm}

\usepackage{booktabs}
\usepackage[format=plain,font=small,labelfont=bf,textfont=up]{caption}
\usepackage{enumitem}
\setlist[enumerate]{itemsep=0.1ex}
\setlist[itemize]{itemsep=0.1ex}
\usepackage[letterpaper,margin=1.05in,bottom=1.65in]{geometry}
\usepackage{graphicx}
\usepackage[hidelinks]{hyperref}
\usepackage{setspace}
\usepackage{float}
\usepackage{caption}
\usepackage{subcaption}

\usepackage{suppmake}

\usepackage[nohints,tight]{minitoc}

\newcommand\mainref\ref
\newcommand\suppref\ref

\makeatletter
\newcommand\tabinput\@@input
\makeatother

\title{\textbf{On the reliability of published findings\\using the regression discontinuity design\\in political science}}

\author[1]{Drew Stommes}
\author[1,2,3,4]{P. M. Aronow}
\author[1,2]{Fredrik S{\"{a}}vje}

\affil[1]{Department of Political Science, Yale University}
\affil[2]{Department of Statistics \& Data Science, Yale University}
\affil[3]{Department of Economics, Yale University}
\affil[4]{Department of Biostatistics, Yale School of Public Health}

\date{\today}

\begin{document}

\makeatletter%
\begin{NoHyper}\gdef\@thefnmark{}\@footnotetext{\hspace{-1em}%
We thank Matias Cattaneo, Alex Coppock, Andrew Harris, Greg Huber, Will Hunt, Austin Jang, Josh Kalla, Winston Lin, John Marshall, Lilla Orr, Tara Slough, Jessica Sun, two anonymous reviewers, and the editor, Abel Brodeur, for helpful comments and discussions.%
}\end{NoHyper}%
\makeatother%

\maketitle

\begin{abstract}
\begin{singlespace}
\noindent%
The regression discontinuity (RD) design offers identification of causal effects under weak assumptions, earning it a position as a standard method in modern political science research.
But identification does not necessarily imply that causal effects can be estimated accurately with limited data.
In this paper, we highlight that estimation under the RD design involves serious statistical challenges and investigate how these challenges manifest themselves in the empirical literature in political science.
We collect all RD-based findings published in top political science journals in the period 2009--2018.
The distribution of published results exhibits pathological features; estimates tend to bunch just above the conventional level of statistical significance.
A reanalysis of all studies with available data suggests that researcher discretion is not a major driver of these features.
However, researchers tend to use inappropriate methods for inference, rendering standard errors artificially small.
A retrospective power analysis reveals that most of these studies were underpowered to detect all but large effects.
The issues we uncover, combined with well-documented selection pressures in academic publishing, cause concern that many published findings using the RD design may be exaggerated.
\end{singlespace}
\end{abstract}

\doparttoc
\faketableofcontents

\clearpage

\section{Introduction}

The regression discontinuity (RD) design requires relatively weak assumptions for identification of causal effects \citep{Cattaneo2020Regression}.
As a result, the method has been widely used in modern empirical political science, and its popularity continues to grow.
Prominent studies examine the effect of incumbency and control over media \citep{Boas2011Controlling}, and the effect of electing extremist candidates in primaries \citep{Hall2015What}.
Because the requisite identifying assumptions are weak, researchers may surmise that all empirical challenges faced when using the method are easily solved.
In this paper, we demonstrate that this view is largely unfounded, highlighting the serious statistical challenges associated with the RD design.

The main contribution of the paper is an investigation of how these statistical challenges manifest themselves in the body of published political science research using the RD design.
We collect all articles using an RD design in the \textit{American Political Science Review}, \textit{American Journal of Political Science}, and \textit{Journal of Politics} published from 2009 through 2018.
We find that the distribution of published RD estimates exhibits some pathological features.
Reported $t$-statistics bunch around, and especially just above, $1.96$, corresponding to the conventional statistical significance level of five percent.
Furthermore, estimated effect sizes are strongly associated with standard error sizes.
This suggests selection pressure to obtain, report or publish results that are statistically significant at conventional levels.

A possible explanation for the pathological features lies with researcher discretion in analysis.
Researcher discretion is an important concern because the RD design offers considerable leeway in how to estimate the effects, and researchers could leverage this leeway to search for significant findings, a practice sometimes called p-hacking.
While the problems of p-hacking and publication bias are widely recognized, their importance for the RD design specifically has largely been overlooked; one recent exception is \citet{Gelman2022Criticism}.
To investigate whether researcher discretion can explain the pathological features, we compare studies that use automated bandwidth selection procedures against those implementing non-automated procedures.
Researcher discretion should be a greater concern with non-automated bandwidths, all else equal, as there are more choices to be made by researchers.
However, we find \emph{more} bunching for studies using automated bandwidth selection, suggesting that researcher discretion in bandwidth selection cannot readily explain our findings.

We next reanalyze all studies with available data in order to investigate the consequences of using inappropriate statistical procedures on the distribution of published results.
Our reanalysis uses a standardized procedure based on current state of the art methods \citep{Calonico2015rdrobust}.
This aims to correct potential methodological shortcomings in studies using older methods and procedures, and to remove discretion in the analysis.
The reanalysis does not meaningfully change the reported point estimates, but the estimated standard errors become larger on average, moving the $t$-statistics closer to zero.
This indicates that the body of published RD studies tends to systematically overestimate the statistical significance of their findings.

A contributing explanation for the pathological features could be selection pressure in the publication process.
Researchers may abandon projects that fail to produce statistically significant results, or journal editors and referees may be reluctant to accept such results for publication.
This type of publication bias would give rise to these pathological features even if researchers do not search through many specifications for statistically significant results.
The problem has been well-documented in, for example, political science \citep{Gerber2008Statistical}, experimental psychology \citep{OSC2015Estimating}, and economics \citep{Brodeur2020Methods}.

The degree to which publication bias is consequential for the health of a literature depends on the statistical power of the analyses conducted in the field.
In a body of underpowered studies, the share of rejected hypotheses that are false positives will typically be much greater than the nominal significance level.
When a field consists primarily of underpowered studies and suffers from severe publication bias, a large majority of published findings could be false.
Inappropriate inference procedures, such as biased standard error estimators, further inflate the number of false positives, exacerbating the problem.
To investigate whether this is a concern for the body of RD studies, we conduct retrospective power analyses on all studies with available data.
The exercise shows that most studies were not well powered to detect small- or moderate-sized effects.
RD studies in political science indeed tend to be underpowered, sometimes severely so, making the concern over possible publication bias more alarming.

The findings in our paper are a stark contrast to the results of a similar study in economics by \citet{Brodeur2020Methods}.
These authors found that the body of RD research in economics does not exhibit notable pathological features, especially compared to studies using instrumental variables.
The difference between political science and economics might be puzzling given that applied empirical practice for RD estimation is similar in the two fields.
However, as noted by \citet[p.\ 3646]{Brodeur2020Methods}, differences between the fields have been documented before.
For instance, \citet{Gerber2008Statistical} find that the ratio of tests just above the $1.96$ threshold of statistical significance relative to those just below is $2$ in political science, while \citet{Brodeur2020Methods} find that the same ratio is only about $1.1$ in economics.
We investigate one potential explanation for the difference regarding RD research: namely that political scientists predominately use election data with the design, which often is a scarce source of data.
Economists use register data or other microdata to a greater degree, which tend to be more plentiful, making the power issues less pronounced.

Taken together, our investigation suggests that many published findings using the RD design in political science may be exaggerated, if not spurious.
Our analysis suggests that lack of sufficient statistical power combined with publication bias is a likely culprit.
Power is a particularly important concern for RD designs in political science as researchers predominantly use the design to study elections, where the amount of data is limited by the number of electoral districts and election cycles.%
\footnote{We thank Tara Slough for suggesting this point to us.}
Although the RD design is an invaluable part of the methodological toolkit in political science, empirical researchers must nevertheless take the estimation challenges associated with the method seriously and properly address them.
Our suggestions for empirical practice moving forward are that researchers should use appropriate, modern analysis procedures, that they should restrict focus to studies with sufficient power, and that they should pre-register analysis plans when possible.

\section{Why is it hard to estimate RD effects?}\label{sec:anatomy-problem}

The causal effect studied in the RD design is the difference between the expected treated and control potential outcomes at a cutoff where treatment assignment changes in a discontinuous fashion.
If the conditional expectation function of the outcome were known, this effect can be calculated without error under an assumption that the conditional expectation functions of the potential outcomes are continuous at the cut point.
In other words, the RD effect is \emph{identified} under a continuity assumption.
The assumption is reasonable in a wide variety of settings, such as when units cannot precisely control treatment assignment.
This has earned the RD design its reputation as a method that produces credible findings.
The concept of an RD design was introduced by \citet{Thistlethwaite1960Regression}.
The formal identification results for the design were first derived by \citet{Hahn2001Identification}.
An insightful and accessible discussion is provided by \citet{Cattaneo2020Practical}.

However, identification is not enough.
The identification exercise presumed the conditional expectation function of the outcome was known, but it will generally not be.
Learning about the causal effect requires estimating this function, and this estimation exercise may be unexpectedly challenging.
While the identifying assumptions required by the RD design might be comparable in strength to those required by randomized controlled trials (RCTs), the \emph{statistical} challenges under the RD design are better compared with those we typically face in other types of observational studies.
The main challenge is that we are interested in the value of the potential outcome functions when evaluated at the cutoff, but there will generally be no observations exactly at the cutoff.
Therefore, researchers need to rely on observations away from the cutoff for estimation, and these observations may not be informative of the quantities of interest.

The standard way to proceed is to assume that the potential outcome functions are smooth, meaning that they do not change too quickly.
This implies that observations close to the cutoff are somewhat informative of the value of the functions at the cutoff, so they can be used for estimation.
Researchers then face a trade-off.
If we estimate the RD effect using only observations very close the cutoff, we ensure that the observations are relevant, but doing so discards most of the data, rendering the estimates less precise, as measured by for example variance.
By contrast, if we include observations farther away from the cutoff, precision improves, but the relevance of the observations decreases.
We can interpret this as a bias--variance trade-off, where the bias is governed by the relevance of the observations and the variance by their numbers.

No matter how one resolves this trade-off, the consequence is that the nominal sample size (i.e., the number of rows in the data set) is not a good measure of the effective sample size (i.e., the amount of useful information in the data set).
Even if identification in an RD design might be almost as credible as in a randomized experiment, we would often need a sample size that is orders of magnitude larger to estimate the causal effect in an RD design to the same level of accuracy as in an experiment.
The bias--variance trade-off also introduces researcher discretion in that there is no intrinsically correct way to resolve the trade-off, and researchers can, purposefully or inadvertently, use this leeway to search for significant findings.

The bias--variance trade-off is inherent to the RD design, and therefore inescapable, but several methods have been developed to help researchers navigate it.
The most prevalent approach is to carefully balance bias and variance with the aim to maximize accuracy, as measured by mean square estimation error.
The bandwidth selection method described by \citet{Imbens2011Optimal} is an early such example.
While this and other similar methods do improve accuracy, they cannot escape the inherent limitations of the data; the effective sample size often remains small even when the nominal sample size is large.
Besides improving accuracy, these methods also reduce researcher discretion, possibly limiting researchers' ability to do specification searching.
Of course, researcher discretion is not entirely removed.
Researchers may, for example, still be able to choose between different bandwidth selection methods, and decide on other aspects of the analysis.

The use of bandwidth selection methods to maximize accuracy provides a partial solution to some of the most pressing problems associated with RD designs, but in doing so, they create a new one.
To conduct hypothesis tests or construct confidence intervals, researchers need to estimate the accuracy of the RD point estimate.
Conventional methods used to gauge this accuracy rely on an assumption that the bias of the estimator is negligible compared to its variance, as would be the case in, for example, a randomized experiment.
However, when a bandwidth selection method balances bias and variance to maximize accuracy, the bias is not negligible, and it must therefore be accounted for when drawing inference.
Researchers sometimes neglect doing so, with the consequence that their hypothesis tests and confidence intervals are misleading even when the sample is large.

There are predominately three approaches to address this concern.
The first approach is so-called undersmoothing, meaning that one uses a bandwidth that is smaller than the optimal one.
If done correctly, this will ensure that the bias indeed is negligible in large samples, so researchers can use conventional methods to construct hypothesis tests and confidence intervals.
The obvious downside is that the accuracy of the point estimator is worse compared to when using the optimal bandwidth.

The second approach is to adjust the inference procedure for the bias.
One can show that the bias when using the optimal bandwidth can be estimated using a similar approach as for the point estimator.
This bias estimate can be used to adjust the point estimate before the construction of confidence intervals and hypothesis tests, allowing researchers to address the bias and still use the optimal bandwidth.
To ensure that the bias adjustment itself does not affect the overall precision, one must ensure that the bias is estimated with considerably higher precision than the precision of the point estimator.
This cannot be ensured in finite samples, but it can be ensured asymptotically if the bandwidth used for the bias adjustment is considerably larger than the bandwidth used for the point estimator.
However, \citet{Calonico2014Robust} point out that it might be undesirable to use a larger bandwidth for the bias adjustment, and that the finite sample behavior might nevertheless be poor.
The alternative approach described by \citet{Calonico2014Robust}, which is becoming the standard approach used in political science, is to also estimate the precision of the bias adjustment and incorporate this into the variance estimator.
This allows researchers to use bias adjustments that have similar precision as the point estimator (e.g., as would be the case when the same bandwidth is used for both point and bias estimation), and it should improve finite sample performance.

The third approach is to adjust the inference procedure with the worse-case bias consistent with the observed data.
The worst-case is taken with respect to a stipulated class of functions, in which the true conditional expectation function is assumed to exist.
This approach obviates the need to estimate the bias, which means that we have to rely on fewer asymptotic approximations and assumptions than with the second approach.
As a consequence, the third approach is expected to perform better in terms of coverage in finite samples, at the cost of typically being more conservative.
The method described by \citet{Armstrong2020Simple} is an example of this approach.

Taken together, the estimation challenges associated with the RD design are immense, comparable to those faced in other types of observational studies.
Yet, being attracted by its weak identification assumptions, researchers may overlook these challenges.
The concerns become even more pressing when the nominal sample sizes are small, as they often are for RD studies in political science.
In particular, the most common type of RD study in political science uses election cutoffs to study the effect of incumbency or other characteristics of candidates or parties.
The sample size is here limited by the number of election districts and the frequency of the elections, which both tend to be small.
For example, if we are interested in contemporary American gubernatorial elections, we are limited to around 200 observations, of which most will not be close elections and thus not relevant when estimating the RD effect.
These estimation challenges and the relatively limited number of observations in election RDs, in particular, motivate us to take a closer look at the state of RD research in political science.

\section{The State of RD Studies in Political Science}\label{sec:state-of-lit}

We collected all studies using an RD design published in the \textit{American Political Science Review}, \textit{American Journal of Political Science}, and \textit{Journal of Politics} between 2009 and 2018.
We included studies implementing an RD design as the primary empirical strategy in an applied setting, in addition to studies where an RD design complements another design \citep[e.g.,][]{Broockman2016Preaching}.
We excluded studies primarily making methodological contributions \citep[e.g.,][]{Cattaneo2016Interpreting}.
There were 45 studies in total that satisfied these inclusion criteria.
Section~\suppref{sec:sample-selection} in the supplement describes the sample selection strategy in more detail and contains a list of all 45 studies.

\begin{figure}
	\centering
	\captionsetup[subfigure]{justification=centering}
	\centering
	\begin{subfigure}{.40\textwidth}
		\centering
		\includegraphics[width=.99\linewidth]{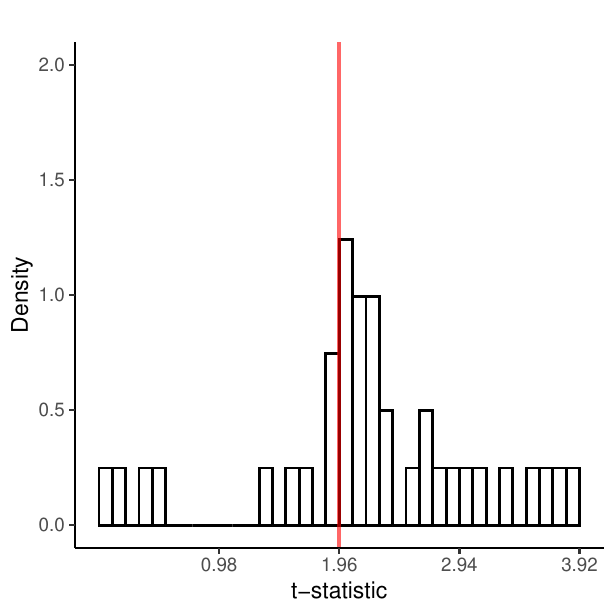}
		\caption{\footnotesize{Main reported result}}
		\label{fig:orig_tscore_hist_one_est}
	\end{subfigure}%
	\hspace{0.2in}
	\begin{subfigure}{.40\textwidth}
		\centering
		\includegraphics[width=.99\linewidth]{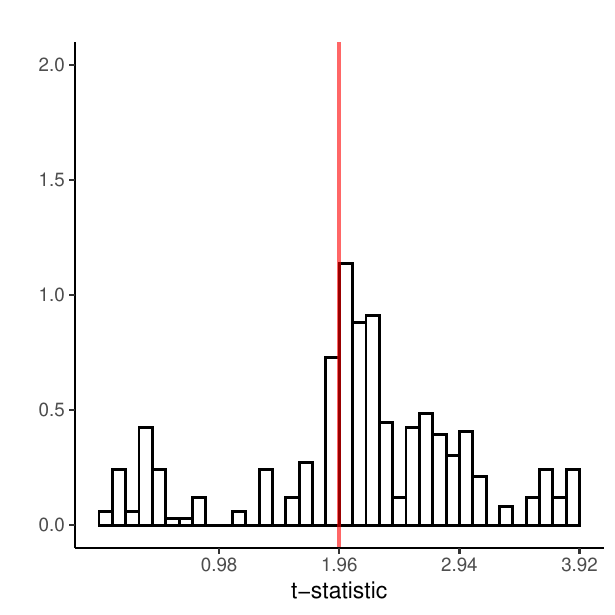}
		\caption{\footnotesize{All primary results}}
		\label{fig:orig_tscore_histogram}
	\end{subfigure}
	\caption{Distribution of $t$-statistics among published RD studies in political science}
	\label{fig:orig_tscores}
\end{figure}

Figure~\ref{fig:orig_tscores} presents the distribution of $t$-statistics for the published findings among all 45 studies in our sample.
The first panel presents the distribution of the main reported result of each article, meaning that each article contributes one $t$-statistic.
The second panel presents the distribution of all results that were referenced in the articles' abstracts, which we take as a proxy for being a primary result of an article.
There are $80$ $t$-statistics in total in the second panel, meaning that each article contributes $1.78$ statistics on average.
To avoid overrepresentation of articles that present many results, the $t$-statistics in the second panel are weighted by the inverse of the total number of results in each study.
For the majority of the studies, we either use reported $t$-statistics or reconstruct $t$-statistics using the ratio of coefficients and standard errors.
For the small number of the studies where neither $t$-statistics nor standard errors were reported, we reconstruct $t$-statistics using reported $p$-values, confidence intervals and point estimates, or using replication code and data provided by the authors.

The histograms in both panels of Figure~\ref{fig:orig_tscores} demonstrate clustering around the value $1.96$.
This corresponds to a significance level of $5\%$, which is the conventional threshold for calling a result ``significant.''
Particularly noteworthy is the substantial imbalance in density to the right of $1.96$, suggesting that results that clear the $5\%$ significance level threshold are artificially favored.
The density of $t$-statistics increases somewhat before the $1.96$ threshold, which could indicate that almost significant results are also favored.
Nevertheless, the spike in density to the right of $1.96$ is pathological, in the sense that we would not expect to observe this pattern of $t$-statistics occurring naturally.

The number of RD studies in our population is relatively small, and it is perhaps conceivable that this distribution of $t$-statistics could arise with somewhat high probability even when there are no real pathologies in the literature.
To assess this in a more formal way, we turn to the so-called caliper test described by \citet{Gerber2008Statistical,gerber-malhotra-2008}, which was also used by \citet{Brodeur2020Methods}.%
\footnote{We thank one of the reviewers and the editor for the suggestion of adding these caliper tests.}
The test compares the number of studies just above and just below the $1.96$ threshold that corresponds to statistical significance at the $5\%$ level.
Under the assumption (null hypothesis) that the $5\%$ significance level is irrelevant for researchers' and editors' publication decisions, the $1.96$ threshold should be immaterial, and we would expect approximately the same number of studies just above and just below it.
However, the $1.96$ threshold becomes relevant for the publication decision if there is p-hacking or publication bias.
In this case, we would not surprised to see more studies just above the threshold.
Following \citet{Gerber2008Statistical,gerber-malhotra-2008}, we include studies within windows around the cutoff with radii (i.e., half-widths or calipers) of $10\%$, $15\%$ and $20\%$ of the threshold itself, here corresponding to $0.196$, $0.294$ and $0.392$, respectively.
A smaller window is better targeted towards the $1.96$ threshold, but has lower power.
To account for the likely dependence between estimates from the same study, we include only the main estimate from each study in this test.

Table~\ref{tab:caliper-all} presents the results from the caliper tests.
For all three window sizes, there are approximately three times as many studies just above the threshold than just below it.
We calculate the probability of observing an imbalance at least this extreme under the null hypothesis of no real pathologies in the RD literature by using a binomial distribution with equal success probability, capturing that we would expect a study to be equally likely to be just above and below the threshold under the null.
Following \citet{Gerber2008Statistical,gerber-malhotra-2008}, the test is one-tailed, reflecting the fact that we feel comfortable ruling out, before seeing the data, that researchers or editors actively favor insignificant findings.
The $p$-values of the caliper tests range between $3\%$ and $7\%$, depending on the window size, indicating that it is unlikely that we would observe this large of an imbalance in the $t$-statistics if there were no real pathologies in the RD literature.
These findings are in line with \citet{Gerber2008Statistical,gerber-malhotra-2008}.
However, as noted in the introduction, the findings are in contrast to the findings of \citet{Brodeur2020Methods}, who document fewer pathological features for the RD design in economics than other commonly used methods.

\begin{table}[ht]
\caption{Caliper tests of imbalance of $t$-statistic among RD studies}\label{tab:caliper-all}
\centering
\begin{tabular}{llll}
	\tabinput{tables/table1-main-caliper-test.tex}
\end{tabular}
\end{table}

\section{Researcher discretion}\label{sec:researcher-disc}

A possible explanation for the bunching of $t$-statistics around the value $1.96$ observed in the previous section is researcher discretion.
Such discretion could be used to seek statistically significant results by searching through specifications and analysis procedures.
When done deliberately, the practice is called p-hacking, but researchers can conduct specification searching inadvertently.
No matter whether it is deliberate or not, searching for significant results invalidates hypothesis tests, $p$-values and confidence intervals.

It is difficult to directly gauge the influence of researcher discretion and possible p-hacking.
We do not observe all specifications researchers tried before finding the one reported in their published article, and unusual specification choices can often be rationalized after the fact.
In an effort to circumvent this problem, we take advantage of the fact that different methods of bandwidth selection provide different levels of researcher discretion.
An automated bandwidth selection procedure, such as those described by \citet{Imbens2011Optimal} and \citet{Calonico2014Robust}, leaves less room for specification searching.
This is in contrast to non-automated procedures and global polynomial specifications, which involve many specification choices.
Table~\suppref{tab:rd_type_table} in the supplement describes the bandwidth selection procedures used by the studies in our investigation.
This table shows that while automated procedures have seen increased use over time, roughly half of all studies throughout the studied period use non-automated procedures.

\begin{figure}
	\centering
	\captionsetup[subfigure]{justification=centering}
	\centering
	\begin{subfigure}{.40\textwidth}
		\centering
		\includegraphics[width=.99\linewidth]{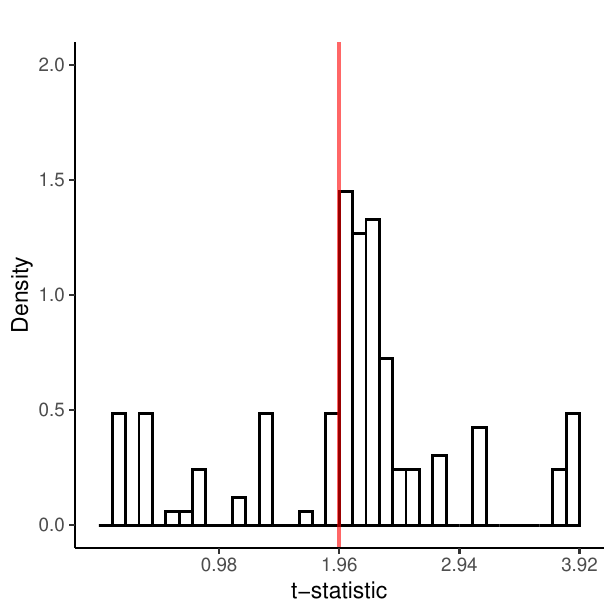}
		\caption{\footnotesize{Automated}}
		\label{fig:orig_allauto_tscore_histogram}
	\end{subfigure}%
	\hspace{0.2in}
	\begin{subfigure}{.40\textwidth}
		\centering
		\includegraphics[width=.99\linewidth]{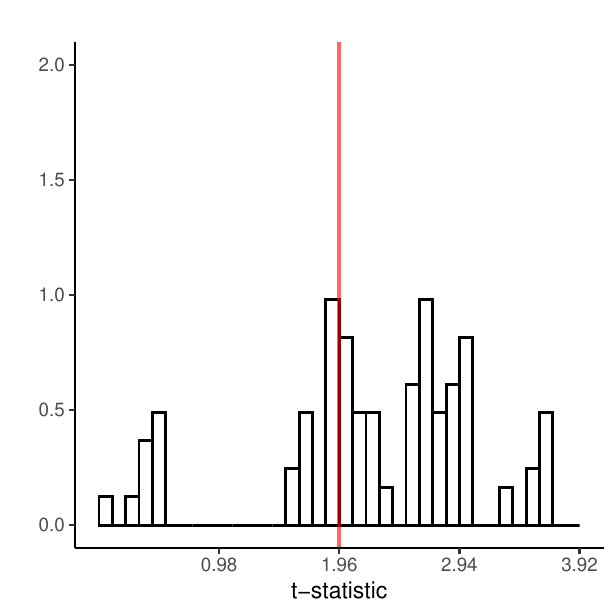}
		\caption{\footnotesize{Non-automated}}
		\label{fig:orig_allnotauto_tscore_histogram}
	\end{subfigure}
	\caption{Distributions of $t$-statistics disaggregated by bandwidth selection procedure}
	\label{fig:orig_tscores_bw_comparison}
\end{figure}

Figure~\ref{fig:orig_tscores_bw_comparison} presents histograms of $t$-statistics disaggregated by studies that use and do not use an automated bandwidth selection procedure.
The first panel consists of studies using automated methods, and we see a clear spike to the right of $1.96$, similar to the finding in the previous section.
The second panel, which consists of studies using non-automated methods, does not present a clear spike, although there are more studies to the right than to the left of the cutoff.
To the degree that the bandwidth selection method is a good proxy of researcher discretion, this finding suggests that discretion is not the driving force of the pathological features we observed in the previous section.

We conduct the caliper tests described in the previous section in the two groups of studies using automated and non-automated bandwidth selection procedures.
The results are presented in Section~\suppref{sec:additional-caliper} of the supplement.
In summary, there are almost the same number of $t$-statistics just above as below the $1.96$ threshold within all windows for the studies using non-automated procedures, with $p$-values around $50\%$, indicating that this is not an unusual observation under the null.
However, there is only one $t$-statistic just below the threshold for studies using automated procedures, irrespectively of window size, and there are 6 to 10 $t$-statistics just above the threshold, depending on window size.
The $p$-values are between $1\%$ and $6\%$, indicating that this would be an unusual observation under the null.
This corroborates the insights given by the histograms.
However, note that the samples are smaller in these disaggregate tests, so power is reduced.

These results should be interpreted with caution.
When researchers use non-automated bandwidth selection methods, they are generally expected to report the estimates using several different bandwidths, with the expectation that the estimates are similar for most of those bandwidths.
This limits the scope of specification searching even when using a non-automated method, making the bandwidth selection method less useful as a proxy for researcher discretion.
There are also other sources of researcher discretion.
For example, unless researchers have pre-committed to a bandwidth selection method before having access to the data, the decision to use a particular automated bandwidth selection method may itself be part of the specification search.
Researchers also have discretion over which questions to investigate, and they may search through a large set of different outcomes, trying to find one that is statistically significant at conventional levels.
While the current investigation does not allow us to rule out all types of researcher discretion as an explanation, it suggests that specification searching using bandwidth selection is not a major driving force of the pathological features we saw in the previous section.

\section{Reanalysis}

We next seek to understand what these studies, all else equal, might have looked like had they all been analyzed with the same, modern approach.
This serves two purposes.
First, it provides additional insights about whether researcher discretion is a contributing factor to the pathological features.
By reanalyzing all studies using a standardized procedure, we remove many sources of researcher discretion.
Second, some studies used procedures that do not properly address the statistical concerns outlined in Section~\ref{sec:anatomy-problem}, potentially providing misleading results.
A reanalysis addresses many, if not all, of these statistical concerns.

\subsection{Methodology}

Replication data are required to conduct this reanalysis.
We were able to obtain such data for $36$ out of the $45$ studies in our sample.
Section~\suppref{sec:replication-data} in the supplement describes how we collected the replication data and lists the studies for which we failed to obtain such data, including brief descriptions of why we failed to do so.
The most common reason was the use of proprietary data, which the original authors were prohibited from sharing.

Our primary reanalysis uses the default settings in the R package \texttt{rdrobust} \citep{Calonico2015rdrobust}. The procedure implemented in this package has well-understood theoretical properties and attempts to address the concerns discussed in Section~\ref{sec:anatomy-problem}.
The reanalysis results we present in the main body of the paper exclude all studies that originally used the \texttt{rdrobust} package to estimate the effects, as a re-analysis would bring few new insights for these studies.
This results in $39$ estimates across $25$ studies.
Thus, the investigation presented in this section focuses on these $25$ studies.
The re-analysis for the full set of $36$ studies with replication data is presented in Section~\suppref{sec:all-studies} in the supplement.

One limitation of the approach by \citet{Calonico2015rdrobust} is that it is motivated by an asymptotic approximation.
This means that confidence intervals and hypothesis tests may not be valid in finite samples.
For example, \citet{wagernoisebased} reports Monte Carlo simulations suggesting that even when $n=1,000$, coverage for nominal 95\% confidence intervals can be less than 92\%.
For this reason, we also conduct reanalyses using the \texttt{rdhonest} procedure by \citet{Armstrong2020Simple}, for which asymptotic approximations feature less heavily, as discussed in Section~\ref{sec:anatomy-problem}.
However, \texttt{rdhonest} also has limitations.
Unlike \texttt{rdrobust} which adapts to the degree of smoothness in the regression function, \texttt{rdhonest} requires that the researcher specifies, or estimates, a lower bound on smoothness.
In our setting, we use the global quartic regression implemented in \texttt{rdhonest} to estimate a bound on smoothness, which requires stronger assumptions than if the smoothness bound was selected based on prior knowledge.

\subsection{Results}

\begin{figure}
	\centering
	\captionsetup[subfigure]{justification=centering}
	\centering
	\begin{subfigure}{.33\textwidth}
		\centering
		\includegraphics[width=.99\linewidth]{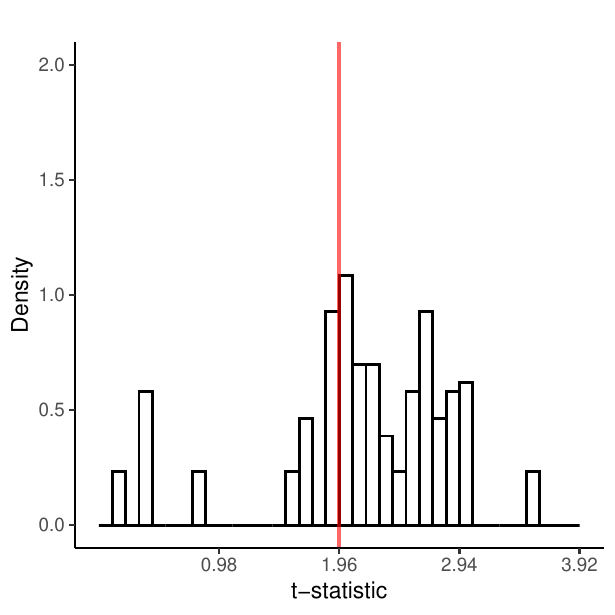}
		\caption{Original approach}
		\label{fig:orig_tstat_noncctrean}
	\end{subfigure}%
	\begin{subfigure}{.33\textwidth}
		\centering
		\includegraphics[width=.99\linewidth]{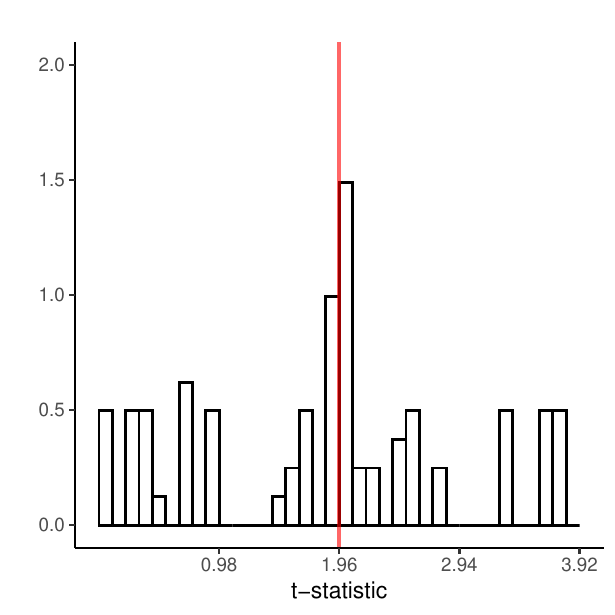}
		\caption{CCT bandwidth}
		\label{fig:classical_tstat_noncctrean}
	\end{subfigure}%
	\begin{subfigure}{.33\textwidth}
		\centering
		\includegraphics[width=.99\linewidth]{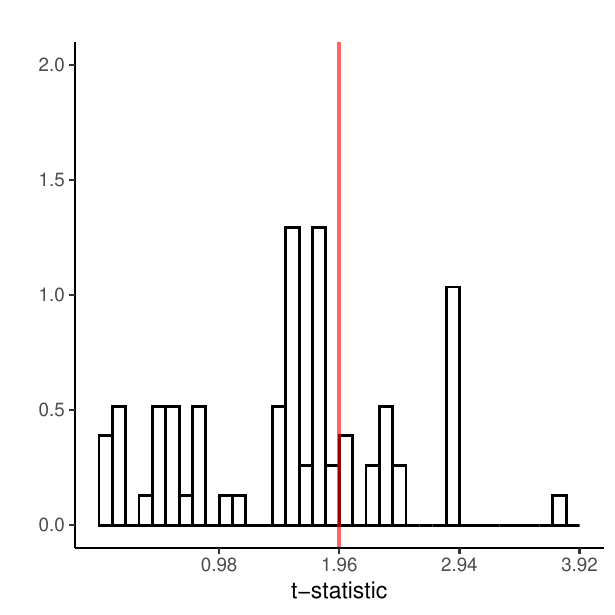}
		\caption{Bias corr.\ and robust SEs}
		\label{fig:cct_tstat_noncctrean}
	\end{subfigure}
	\caption{Distributions of $t$-statistics among replicated studies by method of analysis}
	\label{fig:tstat_noncctreanalyzed}
\end{figure}

Figure~\ref{fig:tstat_noncctreanalyzed} presents the distribution of $t$-statistics among the $39$ estimates from the $25$ studies included in the main replication exercise.
Panel~\ref{fig:orig_tstat_noncctrean} replicates Figure~\ref{fig:orig_tscore_histogram} for the $39$ estimates, containing the reported $t$-statistics using the analysis procedures used in the original studies.

Panel~\ref{fig:classical_tstat_noncctrean} re-analyzes all $39$ estimates using the default bandwidth selection method implemented in \texttt{rdrobust} (``mserd'') but does not make any adjustments for bias for the point estimator nor any adjustments for the standard errors.
For this reason, these $t$-statistics will be too large as the accuracy of estimates is overestimated, potentially making them misleading.
However, if specification searching in bandwidth selection was an important driver, there would be noticeable differences between the first and second panels.
The figure demonstrates that the distribution shifts to the left compared to the first panel, but a clear spike remains at the value $1.96$ in the second panel.
This provides additional support to the conclusion that specification searching over bandwidths is not the main driver of the findings.

Panel~\ref{fig:cct_tstat_noncctrean} re-analyzes the studies using the same bandwidth selection method as in the second panel but also implements the bias adjustment for the point estimator and adjusts the estimated standard errors to account for the added imprecision of the bias adjustment.
Here, we see a clear shift to the left, indicating that these studies have systematically overestimated the accuracy of their estimates.

\begin{figure}
	\centering
	\captionsetup[subfigure]{justification=centering}
	\centering
	\begin{subfigure}{.33\textwidth}
		\centering
		\includegraphics[width=.99\linewidth]{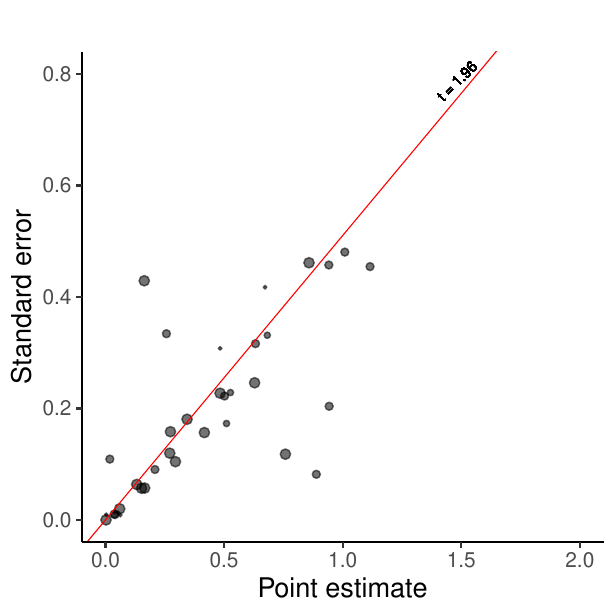}
		\caption{Original approach}
		\label{fig:orig_funnel_noncctrean}
	\end{subfigure}%
	\begin{subfigure}{.33\textwidth}
		\centering
		\includegraphics[width=.99\linewidth]{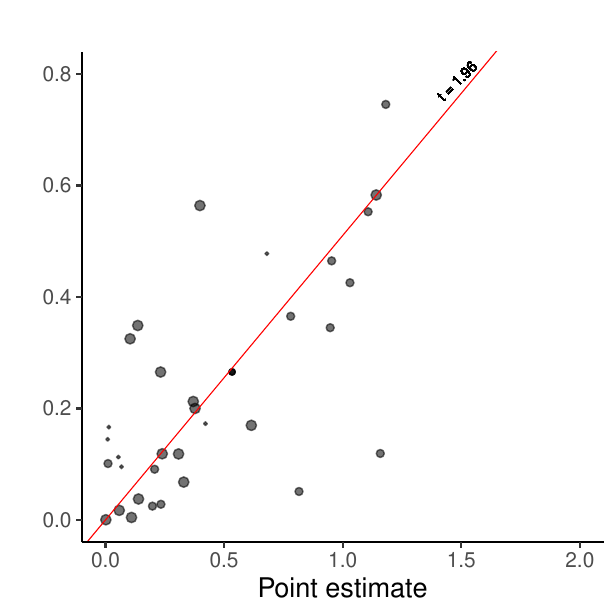}
		\caption{CCT bandwidth}
		\label{fig:classical_funnel_noncctrean}
	\end{subfigure}%
	\begin{subfigure}{.33\textwidth}
		\centering
		\includegraphics[width=.99\linewidth]{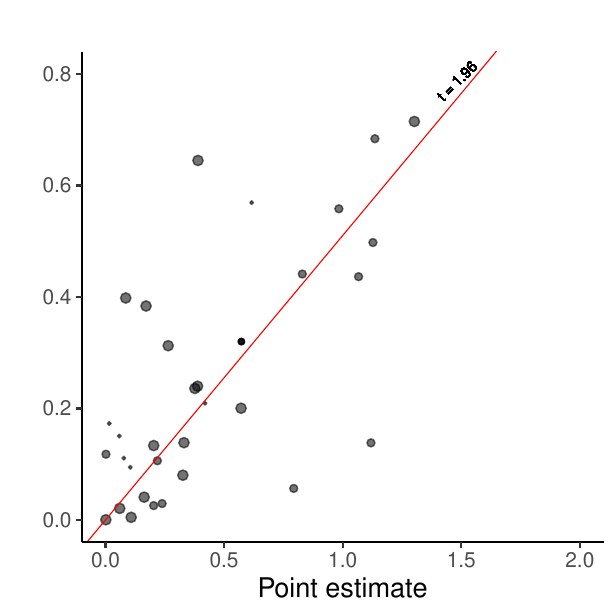}
		\caption{Bias corr.\ and robust SEs}
		\label{fig:cct_funnel_noncctrean}
	\end{subfigure}
	\caption{Funnel plots for replicated studies by method of analysis}
	\label{fig:funnel_plots_noncctreanalyzed}
\end{figure}

We can disaggregate the information in Figure~\ref{fig:tstat_noncctreanalyzed} using a funnel plot, which is a scatter plot with standardized point estimates and standardized estimated standard errors on the two axes.
Figure~\ref{fig:funnel_plots_noncctreanalyzed} presents such a funnel plot for the $39$ estimates in the replication study.%
\footnote{Because the funnel plots require standardization, we cannot construct these plots for studies without replication data, which is why they were not used in Section~\ref{sec:state-of-lit}.}
We use the sample variance of the outcome among control units for the standardization.
The boundary for statistical significance at the $5\%$ level is a line emanating from the origin with a slope of $1.96$, here drawn in red, where estimates to the right of this line attain statistical significance at the conventional level.
Like in Figure~\ref{fig:tstat_noncctreanalyzed}, the first panel presents the estimates from the original studies, the second panel uses the default bandwidth selection method in \texttt{rdrobust} but no bias adjustment, and the third panel implements the bias adjustment.

We find that the re-analysis does not systematically change point estimates, but it does have meaningful consequences for the estimated standard errors.
The standard errors are systematically larger in Figure~\ref{fig:classical_funnel_noncctrean} than in Figure~\ref{fig:orig_funnel_noncctrean}.
This is partly because some studies use global polynomial specifications to estimate the RD effect.
A global polynomial specification does not impose any bandwidth restriction on the sample, making the bandwidth selected by \texttt{rdrobust} dramatically smaller, leading to larger standard errors.
The standard errors are even larger in Figure~\ref{fig:cct_funnel_noncctrean}, reflecting the added imprecision introduced by the bias adjustment.

We note that the estimates cluster around the red line corresponding to a $t$-statistic of $1.96$, especially in the first panel.
Echoing the logic of \citet{Gerber2001Testing}, it is hard to rationalize such a strong positive correlation between the point estimates and standard errors.
In theory, one explanation could be that researchers are tremendously adept at power analysis, so they are able to target the sample size to have just sufficient power to detect the effect of interest.
However, this explanation appears improbable; there are many uncertainties involved with a power analysis, and researchers rarely have precise control over the sample size in an RD study.
A more probable explanation is selection pressure on what type of results get reported and published.

The results of the reanalysis are largely unchanged when we use the \texttt{rdhonest} procedure by \citet{Armstrong2020Simple} in place of \texttt{rdrobust}.
Section~S8 in the supplement reports the results using this alternative approach.
The version of the \texttt{rdhonest} package we used for the re-analysis cannot account for statistical clustering, so our replication using this package is restricted to studies that did not originally use clustered standard errors.
Because the \texttt{rdhonest} package uses a worst-case bias adjustment, conventional $t$-statistics will not capture the adjustment made here.
To make the results from \texttt{rdhonest} comparable to the other results in this paper, we derive pseudo $t$-statistics based on the width of the confidence intervals produced by the \texttt{rdhonest} package.
Substantively, the distribution of reanalysis results is similar to those produced by \texttt{rdrobust}: the pseudo $t$-statistics shift to the left, and the pseudo standard errors systematically increase.

\section{Power Analysis}

Having well-powered studies is one of the best defenses against the concerns highlighted in this paper.
Selection pressure for significant findings in a body of poorly powered studies could make most of the reported results spurious.
This is because the probability of rejecting a false null hypothesis is not much greater than rejecting a true null hypothesis when power is low, meaning that the proportion of false positives of all positive findings in the published literature will be approximately the same as the proportion of true null hypothesis in the body of conducted studies.
In some literatures, most investigated (alternative) hypotheses are false, meaning that most published positive findings also will be false if the power is low.
But in a body of well-powered studies, the true positives will be a disproportionally large share of the overall positive findings, decreasing the proportion of published false positives.
Furthermore, a well-powered study can in some cases make researcher discretion less consequential, because there is less variability in the estimate to exploit.
Therefore, as noted in Section~\ref{sec:anatomy-problem}, it is worrying that the RD design is unusually demanding with respect to sample size, and even samples that on the surface appear large can be poorly powered.

\begin{table}
	\footnotesize
	\setlength{\tabcolsep}{10pt}
	\centering
	\caption{Proportion of analyses achieving 60\%, 80\% and 95\% power by effect size}\label{tab:power_results}
	\begin{tabular}{c ccc}
		\tabinput{tables/table2-power-results.tex}
	\end{tabular}
\end{table}

To investigate the extent to which this is a relevant concern among RD studies in political science, we conducted retrospective power analyses for all $36$ studies with replication data, comprising $64$ separate analyses.
Using the power analysis method implemented in the R package \texttt{rdpower} by \citet{Cattaneo2019Power}, we estimate the power of a two-tailed test at the $5\%$ significance level based on a central limit approximation for the sampling distribution.
The power analyses presume that $p$-values will be constructed with the bias adjustment and robust standard errors as implemented by \texttt{rdrobust}.
As this procedure is only asymptotically valid, our power calculations may be optimistic relative to approaches that control the false positive rate in finite samples.

We investigate power with respect to four difference effect sizes, ranging from small to large effects.
We measure effect size by Cohen's $d$, which is the treatment effect standardized by the standard deviation of the outcome \citep{Cohen1988Statistical}.
We use the standard deviation of the outcome of control units within the default bandwidth in the \texttt{rdrobust} package for this standardization.
We investigate the effect sizes $0.1$, $0.2$, $0.5$ and $0.8$.
While $0.5$ is conventionally labelled as a medium-sized effect, modern social science tends to study effects of smaller sizes.
For example, the What Works Clearinghouse, which is a governmental program that collects and reviews evidence of the effectiveness of various policies, labels an effect size of $0.25$ as ``substantively important'' in their handbook \citep{WWC2017Procedures}.

Table~\ref{tab:power_results} presents the results from the power analyses.
The cells give the proportion of the $64$ analyses that attain the power specified by the columns for the effect size specified by the rows.
For example, we see that only $17\%$ of the analyses attain $60\%$ power to detect a standardized effect of $0.1$.
We weigh the analyses in the same way as in Figure~\ref{fig:orig_tscore_histogram} to account for the fact that some studies conducted more analyses than others.

The table shows that these studies are overall poorly powered to detect anything but large effects.
Only $22\%$ of the analyses achieve $80\%$ power to detect a $0.2$ effect size.
Recall that $80\%$ power is the conventional level that researchers often strive to achieve when designing a study.
Hence, about four out of five RD studies are not properly powered for effect sizes that in many applications would be considered noteworthy.
For the effect sizes of $0.5$ and $0.8$, the shares of properly powered studies increase to $56\%$ and $70\%$, respectively.
Both $0.5$ and $0.8$ would be very large effects in most modern literatures in political science, so the fact that almost half of the studies are not well-positioned to detect such effects gives an indication of the severity of the problem.
Overall, the body of RD studies is alarmingly underpowered.

\begin{table}
	\footnotesize
	\setlength{\tabcolsep}{10pt}
	\centering
	\caption{Proportion of estimates achieving 60\%, 80\% and 95\% power by effect size}\label{tab:power_results_elec}
	\begin{subtable}{.40\textwidth}
		\centering
		\caption{\footnotesize{Studies Using Elections}}
		\begin{tabular}{c ccc}
		\tabinput{tables/table3a-election-power-results.tex}
		\end{tabular}
	\end{subtable}%
	\hspace{0.2in}
	\begin{subtable}{.40\textwidth}
		\centering
		\caption{\footnotesize{Studies Not Using Elections}}
		\begin{tabular}{c ccc}
			\tabinput{tables/table3b-nonelection-power-results.tex}
		\end{tabular}
	\end{subtable}
\end{table}

As previously discussed, one possible reason for the lack of power in RD studies in political science lies with the fact that many RD studies in the field use data from elections.
The number of elections in a given region is necessarily finite, often with no way to accumulate more data of a similar type through additional data collection efforts.
We investigate this as a potential explanation by examining the power of such studies.
Of the studies in our investigation, 68.9\% use electoral data, and among studies with replication data, 72.2\% use electoral data.
Table~\ref{tab:power_results_elec} presents power for the studies disaggregated by data source.
We see that studies in political science that use electoral data have considerably lower power than those that do not.
Along with general differences in disciplinary norms in replicability across fields, this helps to explain why our results are more pessimistic for the RD design than those of \citet{Brodeur2020Methods} in the field of economics.

\begin{figure}
	\centering
	\captionsetup[subfigure]{justification=centering}
	\centering
	\begin{subfigure}{.40\textwidth}
		\centering
		\includegraphics[width=.99\linewidth]{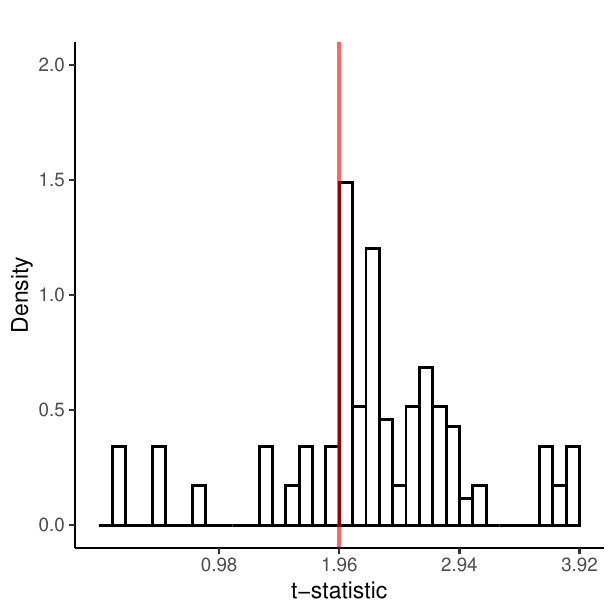}
		\caption{\footnotesize{Election data}}
	\end{subfigure}%
	\hspace{0.2in}
	\begin{subfigure}{.40\textwidth}
		\centering
		\includegraphics[width=.99\linewidth]{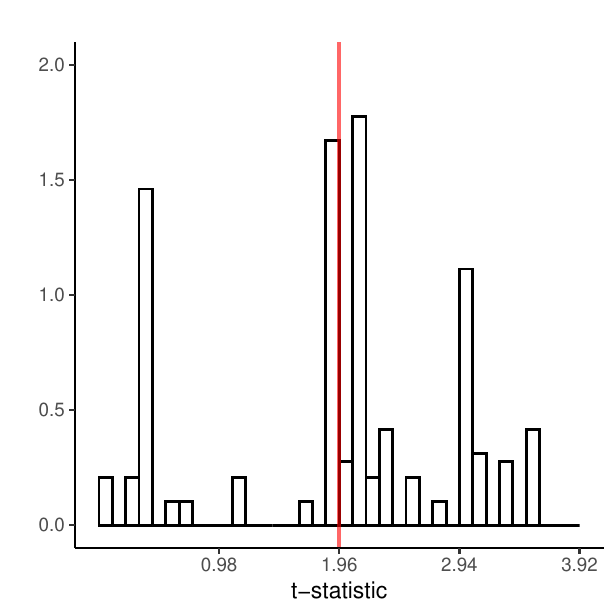}
		\caption{\footnotesize{All other data sources}}
	\end{subfigure}
	\caption{Distributions of $t$-statistics disaggregated by data source}
	\label{fig:tscores-by-data-source}
\end{figure}

To get a sense of how consequential these power issues are for the pathological features we observed in Section~\ref{sec:state-of-lit}, we consider the distributions of $t$-statistics separately for studies using election data and those that do not, in a similar fashion as the investigation in Section~\ref{sec:researcher-disc}.
The distributions are presented in Figure~\ref{fig:tscores-by-data-source}.
We see that there are noticeably more $t$-statistics just above the $1.96$ threshold for studies using election data, while the $t$-statistics are more balanced around the threshold for the other studies.
Although, there is a noticeable spike at the threshold for the second set of studies as well.
This observation is consistent with insufficient statistical power being the main driver of the pathological features, but this investigation is far from conclusive.
It could, for example, be that election research is more competitive in political science, meaning that researchers have a greater incentive to p-hack and that editors are more selective.
Additionally, we are limited in the conclusions we draw about studies using non-election data given the limited number of published papers using such data.

While power is a particularly important consideration for the RD design, given its hunger for data, the issues we document here are not unique to RD studies.
\citet{ArelBundock2022Quantitative} use meta-analytic evidence to estimate that approximately 1 in 10 tests reported in political science have at least 80\% power.
This analysis relies on the assumption that studies in a given literature share a common mean effect (inclusive of bias), which may understate the power of many studies in the presence of effect heterogeneity or differential bias across studies.
Nevertheless, the striking degree of the documented issues suggest that power issues are prevalent throughout the field.

\section{Concluding remarks}

The body of published political science research using the RD design exhibits pathological features consistent with selective reporting and publishing based on whether results clear artificial significance thresholds.
More than half of the studies in our sample do not properly estimate the accuracy of their estimates, leading to the associated hypothesis tests and confidence intervals being misleading.
Most of the studies are also underpowered, and are able to detect only large effects.
Taken together, this paints a somber picture of the state of applied studies using the RD design.
Our results suggest that many published findings using the design are exaggerated if not altogether spurious.

The conclusion is that researchers using the RD design must take the estimation challenges associated with the design more seriously.
They should make sure to conduct hypothesis tests and construct confidence intervals in a well-motivated way that properly reflects the accuracy of the finding.
The procedures described by \citet{Calonico2014Robust} and \citet{Armstrong2020Simple} represent significant progress toward this goal.

Furthermore, statistical power will often be a first-order concern with the RD design, and researchers should make sure that they have sufficiently large samples to have a good chance to detect effects of sizes relevant to the question at hand.
When gauging power, researchers should remember that the nominal sample size is not relevant in an RD study, and they should instead consider how much information there is about the potential outcomes close to the cutoff.
Sample size is often beyond the control of the researcher in RD studies; if the accessible sample is too small, researchers should ask whether it is appropriate to conduct the study at all.
The decision to abandon a study due to concerns about power should be taken before running the analysis and observing the estimated effect.

Covariates can sometimes be used to improve precision in an RD study without collecting additional observations, which could alleviate some of the concerns about lack of power with the RD design \citep{Calonico2019Regression}.
However, covariate adjustment involves many largely arbitrary specification choices, which would make it easier to p-hack.
It is unclear if the body of RD studies overall would benefit from making covariate adjustment an expected part of the analysis with the RD design.
Indeed, one of the central benefits of the RD design is exactly that identification does not require background information about the units, other than the running variable that decides treatment assignment.

Journals should consider taking power into account when making publication decisions.
If a study is severely underpowered, it may not contribute much to the literature even if it has nominally statistically significant results.
Similarly, a well-powered study often provides useful insights no matter if its results are statistically significant at conventional levels, because confidence intervals will typically be narrow.
Being more mindful about power would alleviate publication bias, making the body of published results more reliable and informative.

Another way to alleviate some of these concerns is preregistration for studies using the RD design.
Given that most RD studies are retrospective, it might be difficult for researchers to credibly show that all analyses were registered before they were conducted.
However, insofar as it as possible, having a full record of analyses that were intended would be helpful for the accumulation of knowledge.
They reveal the absence of findings from the published literature due to the ``file drawer'' problem that disproportionately afflicts insignificant findings, either through the filter of publication or through specification searching.
Scholars evaluating a literature (e.g., in meta-analysis) could use these preregistrations to help to undo distortions introduced through the various filters associated with the production and dissemination of knowledge.
Preregistration is also useful to address inadvertent p-hacking, as researchers are less likely to accidentally or thoughtlessly deviate from their original plans when they are recorded.

\begin{singlespace}
\bibliography{refs-rd-rep}
\end{singlespace}

\clearpage
\startsupp{S}{}{supp}

\hbadness=10000

\clearpage

\section{Sample selection}\label{sec:sample-selection}

\subsection{RD articles included}

We collected our sample of 45 articles in two ways. First, we conducted a targeted search on Google Scholar within the three journals of interest. We searched for any articles mentioning terms which directly refer to the RD design (e.g. ``regression discontinuity'') along with common phrases associated with RD estimators (e.g. ``bandwidth'' and ``cut point''). Our second approach entailed searching for these terms on the journals' websites and reading all articles' abstracts from the past ten years.

We included in our sample articles which used an RD as the primary empirical strategy in an applied setting as well as studies where an RD complements another design (e.g., \citealp{Broockman2016Preaching}). We excluded studies primarily making a methodological contribution (e.g., \citealp{Cattaneo2016Interpreting}) and articles whose research settings are analogous to RD contexts but do not fit the standard definition of an RD. For example, some do not use a continuous running variable (e.g., \citealp{DUNNING:2013aa} and \citealp{Kadt:2017aa}).

We classify as ``primary'' RD point estimates those which are referenced in the articles' abstracts. As such, individual studies include multiple RD point estimates that we include in our sample. The one main estimate of each article was based on which of the primary estimates was most highlighted by the authors, or which one was most central to the authors' key conclusion.

Among our sample of articles, we extracted salient information including the type of estimator used, bandwidth selection procedure (if applicable), number of units, point estimates, standard errors, and when explicitly reported, $p$-values. We list all articles' authors, the publication, and RD type in Table \ref{tab:rd_list}.
These studies span a broad range of substantive topics. For instance, \cite{Fouirnaies:2014aa} use a sharp RD to study the link between incumbency and U.S. Congressional campaign contributions, and \cite{SZAKONYI:2018aa} investigates whether a firm director's electoral victory affects the firm's profitability in the future. \cite{CAVAILLE:2019aa} study how an additional year of schooling affects individuals' attitudes towards immigration, and \cite{Boas:2014aa} investigate how certain electoral outcomes affect government contracts.

\begin{singlespace}
	{\renewcommand{\arraystretch}{1.1}
		\begin{table}[H]
			\footnotesize
			\setlength{\tabcolsep}{8pt}
			\centering
			\caption{Sample of Political Science RD Studies/Outcomes}
			\smallskip
			\resizebox{0.80\textwidth}{!}{%
			\begin{tabular}{p{9.5cm}p{1.25cm}p{1.75cm}}
				\toprule
				Author(s) \& Year & Journal & RD type \\
				\midrule
				\cite{Ariga:2015aa} & \emph{JoP} & Sharp  \\
				\cite{Boas:2014aa} & \emph{JoP} & Sharp \\
				\cite{Boas2011Controlling} & \emph{AJPS} & Sharp \\
				\cite{Bohlken:2018aa} & \emph{AJPS} & Sharp \\
				\cite{BROLLO:2012aa} & \emph{APSR} & Sharp \\
				\cite{Broockman2016Preaching} & \emph{AJPS} & Sharp \\
				\cite{Carson:2017aa} & \emph{JoP}  & Sharp \\
				\cite{Caughey:2017ab} & \emph{JoP}  & Sharp \\
				\cite{CAVAILLE:2019aa} & \emph{APSR} & Fuzzy \\
				\cite{CLINTON:2018aa} & \emph{APSR} & Sharp \\
				\cite{Coppock:2016aa} & \emph{AJPS}  & Fuzzy \\
				\cite{DAHLGAARD:2018aa} & \emph{APSR}  & Sharp \\
				\cite{Benedictis-Kessner:2018aa} & \emph{JoP} & Sharp \\
				\cite{Benedictis-Kessner:2016aa} & \emph{JoP} & Sharp \\
				\cite{EGGERS:2009aa} & \emph{APSR} & Sharp \\
				\cite{Eggers:2017ab} & \emph{JoP}  & Sharp \\
				\cite{Erikson:2015aa} & \emph{JoP}  & Sharp \\
				\cite{FERWERDA:2014aa} & \emph{APSR} & Sharp \\
				\cite{FIVA:2018aa} & \emph{APSR} & Sharp \\
				\cite{FOLKE:2016aa} & \emph{APSR} & Sharp \\
				\cite{Folke:2012aa} & \emph{AJPS} & Sharp \\
				\cite{Fouirnaies:2014aa} & \emph{JoP} & Sharp \\
				\cite{GALASSO:2011aa} & \emph{APSR} & Sharp \\
				\cite{Gerber:2011aa} & \emph{JoP} & Sharp \\
				\cite{Gerber:2011ab} & \emph{AJPS} & Sharp \\
				\cite{GULZAR:2017aa} & \emph{APSR} & Sharp \\
				\cite{HAINMUELLER:2017aa} & \emph{APSR} & Fuzzy \\
				\cite{Hall2015What} & \emph{APSR} & Sharp \\
				\cite{HALL:2018aa} & \emph{APSR} & Sharp \\
				\cite{Hidalgo:2016aa} & \emph{AJPS} & Sharp \\
				\cite{Hirano:2011aa} & \emph{JoP} & Sharp \\
				\cite{HOLBEIN:2016ab} & \emph{APSR} & Fuzzy \\
				\cite{Holbein:2016aa} & \emph{AJPS} & Fuzzy \\
				\cite{Klasnja:2015aa} & \emph{JoP} & Sharp \\
				\cite{KLASNJA:2017aa} & \emph{APSR} & Sharp \\
				\cite{LARREGUY:2016aa} & \emph{APSR} & Sharp \\
				\cite{Lopes-da-Fonseca:2017aa} & \emph{AJPS} & Sharp \\
				\cite{MO:2018aa} & \emph{APSR} & Fuzzy \\
				\cite{Novaes:2018aa} & \emph{AJPS} & Sharp \\
				\cite{Palmer:2016aa} & \emph{JoP} & Fuzzy \\
				\cite{Rozenas:2017aa} & \emph{JoP} & Fuzzy \\
				\cite{Sances:2017aa} & \emph{JoP} & Sharp \\
				\cite{Schickler:2010aa} & \emph{JoP}  & Sharp \\
				\cite{SZAKONYI:2018aa} & \emph{APSR} & Sharp \\
				\cite{XU:2015aa} & \emph{APSR} & Sharp \\
				\bottomrule
				\label{tab:rd_list}
			\end{tabular}
		}
		\end{table}
	}
\end{singlespace}

\clearpage

\subsection{Excluded RD articles}

\noindent Some articles tentatively related to the RD design were excluded from our sample.
This was made for primarily two reasons.
First, studies were excluded if their research designs did not meet our definition of an RD design.
For example, we do not consider studies with a clearly discrete running variable as an RD design for the purposes of this paper.
Second, studies were excluded if their primary focus was to develop or refine existing RD estimators, and only used data from previous RD studies as illustration.
We list all these excluded articles below.

\bigskip

\noindent \cite{Anzia:2011aa}:

\noindent The RD-type analysis in their paper is not included as a primary component of their empirical analysis, and while they note that this approach is similar to an RD they do not classify it as one.

\bigskip

\noindent \cite{Cattaneo2016Interpreting}:

\noindent This is a methods-focused paper, deriving an estimator for a certain type of RD design.

\bigskip

\noindent \cite{Caughey:2017aa}:

\noindent This is a methods-focused paper rather than an applied study.

\bigskip

\noindent \cite{CROKE:2016aa}:

\noindent The authors note that their method is ``similar to'' an RD design but not exactly the same because their running variable is not continuous.

\bigskip

\noindent \cite{Kadt:2017aa}:

\noindent Study uses a discrete running variable, which is not consistent with our focus on implementing RD with continuous running variables.

\bigskip

\noindent \cite{DUNNING:2013aa}:

\noindent Despite being inspired by an RD design, it does not quite fit the criteria whereby units are assigned a value on a continuous running variable.

\bigskip

\noindent \cite{Eggers:2015aa}:

\noindent This study mainly focuses on placebo tests and is a broader methodological discussion rather than an application.

\bigskip

\noindent \cite{Eggers:2017aa}:

\noindent This study references different RD studies but does not implement an RD design on its own.

\bigskip

\noindent \cite{Eggers:2018aa}:

\noindent This is a methods-focused paper exploring population-based threshold RD analyses. It is not an applied RD study.

\bigskip

\noindent \cite{FOLKE:2011aa}:

\noindent They employ an additional specification which focuses on close elections and they say is ``similar in spirit'' to RD designs. However, the authors explicitly note how their setting does not allow them to have the same causally identified estimate as an RD and thus do not consider their estimates to be RD estimates.

\bigskip

\noindent \cite{Friedman:2009aa}:

\noindent The analysis does not incorporate the kind of continuous running variable which is standard for RD designs.

\bigskip

\noindent \cite{Gerber:2010aa}:

\noindent This is too far from a standard RD design as it does not employ the typical assumption about continuity at the cut point.

\bigskip

\noindent \cite{Hainmueller:2015aa}:

\noindent The authors are primarily interested in extrapolating RD treatment effects away from the cut point; not a typical RD design.

\bigskip

\noindent \cite{Hopkins:2011aa}:

\noindent Excluded because there were such irregular and major spikes in the density of the running variable that it is behaving more like a discrete variable than a continuous one.

\bigskip

\noindent \cite{Kelley:2015aa}:

\noindent Their setting does not entail the running variable/cut point set-up necessary for a study to qualify as an RD design.

\bigskip

\noindent \cite{Krasno:2008aa}:

\noindent This paper refers to regression discontinuity designs for general motivation for their analysis, which is, in fact, a difference-in-differences estimator.

\bigskip

\noindent \cite{Larreguy:2017aa}:

\noindent The study is primarily interested in an interaction between the RD treatment indicator and a background covariate, which puts the estimator/estimand outside the scope of our focus.

\bigskip

\noindent \cite{Lerman:2017aa}:

\noindent Study uses a discrete running variable, which is not consistent with our focus on implementing RD with continuous running variables.

\bigskip

\noindent \cite{Marshall:2016aa}:

\noindent Study uses a discrete running variable, which is not consistent with our focus on implementing RD with continuous running variables.

\bigskip

\noindent \cite{Mummolo:2018aa}:

\noindent Study uses time as a running variable, and so it's not consistent with our focus on running variables along which units receive a single score to the left or right of a discrete cut point.

\bigskip

\noindent \cite{NELLIS:2018aa}

\noindent This is an RD design combined with an instrumental variables designed in a way that adds several estimation and identification quirks. As such it doesn't neatly fit the definition of an RD study.

\bigskip

\noindent \cite{Nyhan:2017aa}:

\noindent Study uses a discrete running variable, which is not consistent with our focus on implementing RD with continuous running variables.

\bigskip

\noindent \cite{SAMII:2013aa}:

\noindent The study is primarily interested in an interaction between the RD treatment indicator and a background covariate, which puts the estimator/estimand outside the scope of our focus.

\bigskip

\noindent \cite{SEKHON:2012aa}:

\noindent This is a reanalysis of previous results and is methods-centric, rather than an applied paper.

\clearpage

\section{Replication data collection}\label{sec:replication-data}

After generating the list of papers included in our study, we attempted to collect replication data for each one. We used publicly-available data whenever possible (e.g., replication materials made accessible through the Harvard Dataverse repository). Otherwise, we contacted the authors for replication data. We here list all studies for which we could not obtain replication data, along with an explanation of why the data are unavailable:

\bigskip

\noindent \cite{BROLLO:2012aa}:

\noindent The data were not available through an online repository or authors' websites. We contacted the authors multiple times but did not receive a reply.

\bigskip

\noindent \cite{CLINTON:2018aa}:

\noindent Some data necessary for the RD analyses are proprietary and were not available along with the replication materials.

\bigskip

\noindent \cite{FIVA:2018aa}:

\noindent Data required for the RD analyses of interest are government sources and unavailable for public use.
The data are not available online nor from the authors.

\bigskip

\noindent \cite{FOLKE:2016aa}:

\noindent Data required for the RD analyses of interest are government sources and unavailable for public use.
The data are not available online nor from the authors.

\bigskip

\noindent \cite{GALASSO:2011aa}:

\noindent The data were not available through an online repository or authors' websites. We contacted the authors multiple times but did not receive a reply.

\bigskip

\noindent \cite{HAINMUELLER:2017aa}:

\noindent Data required for the RD analyses of interest are government sources and unavailable for public use.
The data are not available online nor from the authors.

\bigskip

\noindent \cite{Hirano:2011aa}:

\noindent Some data necessary for the RD analyses are proprietary and were not available along with the replication materials.

\bigskip

\noindent \cite{HOLBEIN:2016ab}:

\noindent Data required for the RD analyses of interest are government sources unavailable for public use.
The data are not available online nor from the authors.

\bigskip

\noindent \cite{MO:2018aa}:

\noindent Data required for the RD analyses of interest are confidential and unavailable for public use.
The data are not available online nor from the authors.

\clearpage

\section{Yearly number of published RD studies}

\begin{table}[H]
	\footnotesize
	\setlength{\tabcolsep}{10pt}
	\centering
	\caption{Yearly Published RD Articles}\label{tab:rd_journal_breakdown}
	\resizebox{0.98\textwidth}{!}{%
		\begin{tabular}{lccccccccccc}
			\tabinput{tables/tableS2-journal-count.tex}
		\end{tabular}
	}
\end{table}

\begin{table}[H]
	\footnotesize
	\setlength{\tabcolsep}{10pt}
	\centering
	\caption{Bandwidth Selection in Sample of RD Articles}\label{tab:rd_type_table}
	\resizebox{0.98\textwidth}{!}{%
		\begin{tabular}{lccccccccccc}
			\tabinput{tables/tableS3-method-count.tex}
		\end{tabular}
	}
\end{table}

\noindent Note that Erikson et al. (2015) report as primary estimates of interest results derived from different bandwidth selection procedures. Therefore, we count this article twice in Table A3, once for each bandwidth selection, and so Table A3 includes one more unit (46) than Table A2 (45).

\clearpage

\section{Extraction of original t-statistics}\label{sec:coding-tstat}

We collected point estimates, estimated standard errors, and $p$-values whenever they were explicitly reported in the paper. If a study did not report a $p$-value but reports the point estimate and standard error, we derived the implied $p$-value using a normal approximation of the sampling distribution. Some studies did not explicitly report standard errors or $p$-values. We imputed these values as described below. We refer to each article/estimate by the estimate code used in the dataset.

\bigskip

\noindent \cite{Ariga:2015aa} (a-b):

\noindent The standard error estimate was imputed by (a) subtracting the point estimate value from the upper value of the reported 90\% confidence interval, and then (b) dividing that value by the critical value for a 90\% confidence interval (1.64). A two-tailed, normal approximation $p$-value estimate was imputed using point estimate and imputed std. error because $p$-value was not explicitly listed in paper.

\bigskip

\noindent \cite{Caughey:2017ab}:

\noindent The standard error and $p$-value were imputed by running their replication code (which used \texttt{rdrobust}) and extracting the robust standard error value and robust $p$-value.

\bigskip

\noindent \cite{Carson:2017aa}:

\noindent The point estimate was rounded to the nearest tenth in text of original paper; imputed the full estimate that's derived by running the authors' replication code. The standard error and $p$-value were imputed by running their replication code and extracting the values.

\bigskip

\noindent \cite{CLINTON:2018aa} (a-d):

\noindent We imputed the standard error using the width of the confidence interval. We imputed the $p$-value using the reported point estimate and the imputed standard error.

\bigskip

\noindent \cite{Benedictis-Kessner:2018aa} (a-b):

\noindent The standard error was not explicitly listed, and so we imputed it by using the width of the reported confidence interval. The $p$-value explicitly listed in the main body of the article.

\bigskip

\noindent \cite{Benedictis-Kessner:2016aa}:

\noindent The standard error was not explicitly listed and so we imputed it by using the replication files made available for the paper. The $p$-value explicitly listed in the main body of the article.

\bigskip

\noindent \cite{Eggers:2017ab} (a):

\noindent The point estimate was explicitly listed in main body of article, but note that there was a transcription error (the replication yields 1.911 instead of 0.911; the std. error value was transcribed correctly). The standard error explicitly listed in main body of article, and a two-tailed, normal approximation $p$-value estimate was imputed by using the point estimate and std. error.

\bigskip

\noindent \cite{FERWERDA:2014aa}:

\noindent The standard error was imputed by dividing the reported point estimate by the reported t-statistic. A two-tailed, normal approximation $p$-value estimate was imputed by using the reported point estimate and the imputed std. error.
\bigskip

\noindent \cite{HOLBEIN:2016ab} (a-c):

\noindent The standard error was not explicitly listed and so we imputed by using the upper bound of a 95\% CI and the point estimate along with the critical value of 1.96. A two-tailed $p$-value was not explicitly listed; we imputed the two-tailed $p$-value using the imputed SE value along with the point estimate.

\bigskip

\noindent \cite{KLASNJA:2017aa}:

\noindent The standard error was not explicitly listed and so we imputed by using the upper bound of a 95\% CI and the point estimate along with the critical value of 1.96. However, the $p$-value was explicitly listed in main body of article.

\bigskip

\noindent \cite{Schickler:2010aa} (a-d):

\noindent We retrieved the point estimate, standard error estimate, and $p$-value estimate by running the Stata replication files shared by authors.

\clearpage

\section{Additional caliper test results}\label{sec:additional-caliper}

We ran the caliper test separately for studies that implemented an automated bandwidth selection procedure and those that used a non-automated approach to bandwidth selection.
The results in Table~\ref{tab:caliper-auto-bw} presents the imbalance around 1.96 for studies which used automated approaches to bandwidth selection, whereas Table~\ref{tab:caliper-notauto-bw} presents the results for studies which used non-automated approaches.

\begin{table}[H]
	\footnotesize
	\setlength{\tabcolsep}{10pt}
	\centering
	\caption{Caliper tests by bandwidth selection method}
	\begin{subtable}{.40\textwidth}
		\centering
		\caption{Automated BW selection method}\label{tab:caliper-auto-bw}
		\begin{tabular}{llll}
			\tabinput{tables/tableS4a-auto-caliper-test.tex}
		\end{tabular}
	\end{subtable}%
	\hspace{0.5in}
	\begin{subtable}{.40\textwidth}
		\centering
		\caption{Other BW selection method}\label{tab:caliper-notauto-bw}
		\begin{tabular}{llll}
			\tabinput{tables/tableS4b-noauto-caliper-test.tex}
		\end{tabular}
	\end{subtable}
\end{table}

We also ran caliper tests for the t-statistics from the re-analysis.
Table~\ref{tab:cct-conven-caliper-test} presents the results for when the CCT bandwidth selection method is used, corresponding to Figure~\mainref{fig:classical_tstat_noncctrean} in the main paper.
Table~\ref{tab:cct-robust-caliper-test} presents the results for when the CCT bandwidth selection method is used together with the bias adjustment and robust SEs, corresponding to Figure~\mainref{fig:cct_tstat_noncctrean} in the main paper.
Table~\ref{tab:reanalyzed-caliper-test} excludes \citet{DAHLGAARD:2018aa} because their main estimate is a precision weighted estimate that was not part of our re-analysis.

\begin{table}[H]
	\footnotesize
	\setlength{\tabcolsep}{10pt}
	\centering
	\caption{Caliper tests for re-analyzed data}\label{tab:reanalyzed-caliper-test}
	\begin{subtable}{.40\textwidth}
		\centering
		\caption{CCT bandwidth}\label{tab:cct-conven-caliper-test}
		\begin{tabular}{llll}
			\tabinput{tables/tableS5a-cct-conven-caliper-test.tex}
		\end{tabular}
	\end{subtable}%
	\hspace{0.5in}
	\begin{subtable}{.40\textwidth}
		\centering
		\caption{Bias corr. and robust SEs}\label{tab:cct-robust-caliper-test}
		\begin{tabular}{llll}
			\tabinput{tables/tableS5b-cct-robust-caliper-test.tex}
		\end{tabular}
	\end{subtable}
\end{table}

The caliper test is not a data-hungry test, so it will perform quite well with small samples.
However, the samples are worrying small for these additional caliper tests, and power might be a concern.
The results in this section should therefore be interpreted with caution.
The tests were run (and reported) because it was requested during the review process.

\clearpage

\section{Reproduction of originally reported estimates}

This is section we document all failures to reproduce the authors' original estimates during our replication.

\bigskip

\noindent \cite{Benedictis-Kessner:2016aa}:

\noindent
The authors used earlier version of \texttt{rdrobust}. When reproducing their estimates we select the bandwidth through the \texttt{rdbwselect\_2014} for backward compatibility purposes, and the manually feed this bandwidth into the \texttt{rd\-robust} package. The results are quite close but deviate slightly due to the slight differences in estimation code in \texttt{rdrobust} at the time the authors used it and now.

\bigskip

\noindent \cite{EGGERS:2009aa} (a) and (b):

\noindent
There is a slight deviation between our reproduction results and those reported in the paper, though these differences are negligible. The replication data did not include code to allow us to precisely use the same functional form nor the same code to estimate robust standard errors.

\bigskip

\noindent \cite{Erikson:2015aa}  (a) through (c):

\noindent
We received replication files from one author, and while we were able to get quite close with our attempts to reproduce the original estimates, there is a deviation between the ones reproduced and those originally reported in the paper. See STATA code ``presidential elections reg and latex jms.do.''

\bigskip

\noindent \cite{Gerber:2011ab}:

\noindent
The authors noted that they used multiple imputation to fill in missing values for certain variables, though it wasn't exactly clear which ones. Analysis was run on the data that was available. Reanalysis results deviate from those in the paper.

\bigskip

\noindent \cite{XU:2015aa}:

\noindent
We successfully reproduced the loess plots using their STATA code, which did not include code for producing the reported RD estimates. Using the same data, we implemented a loess estimation approach in R which produced a point estimate consistent with that reported in the paper.

\bigskip

\clearpage

\section{P-values}

The following two figures compare $p$-values reported in the original studies and the $p$-values for our re-analysis.
If the original study did not report $p$-values, we have derived implied $p$-values as described in Section~\ref{sec:coding-tstat}.

\begin{figure}[H]
	\centering
	\includegraphics[width=.85\linewidth]{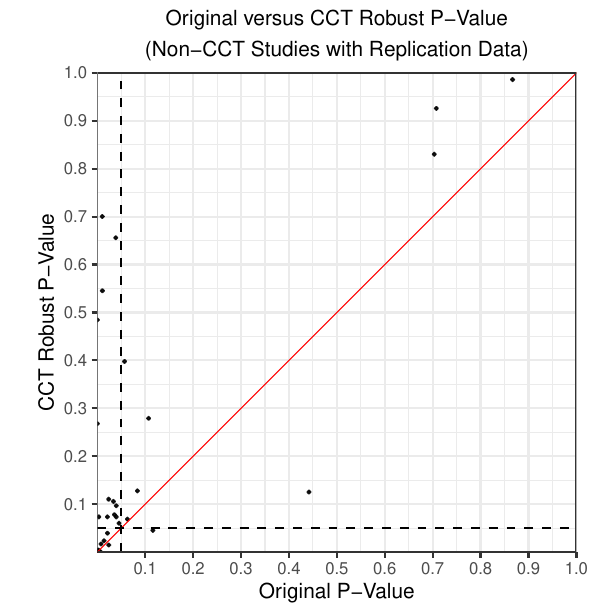}
	\caption{Original vs. CCT $p$-values for non-CCT studies}
	\label{fig:cct_vs_orig_pval_nonCCTs}
\end{figure}

\begin{figure}[H]
	\centering
	\includegraphics[width=.85\linewidth]{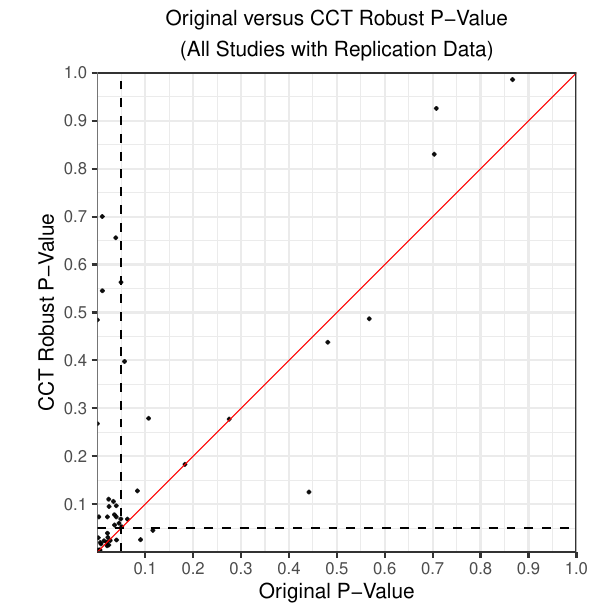}
	\caption{Original vs. CCT $p$-values for all studies}
	\label{fig:cct_vs_orig_pval_all}
\end{figure}

\clearpage

\section{All studies with replication data}\label{sec:all-studies}

The following two figures correspond to Figures~\mainref{fig:tstat_noncctreanalyzed}~and~\mainref{fig:funnel_plots_noncctreanalyzed} in the main paper but includes all studies with replication data.
That is, also studies that used the \texttt{rdrobust} package in their original analysis.

\begin{figure}[H]
	\centering
	\captionsetup[subfigure]{justification=centering}
	\centering
	\begin{subfigure}{.33\textwidth}
		\centering
		\includegraphics[width=.99\linewidth]{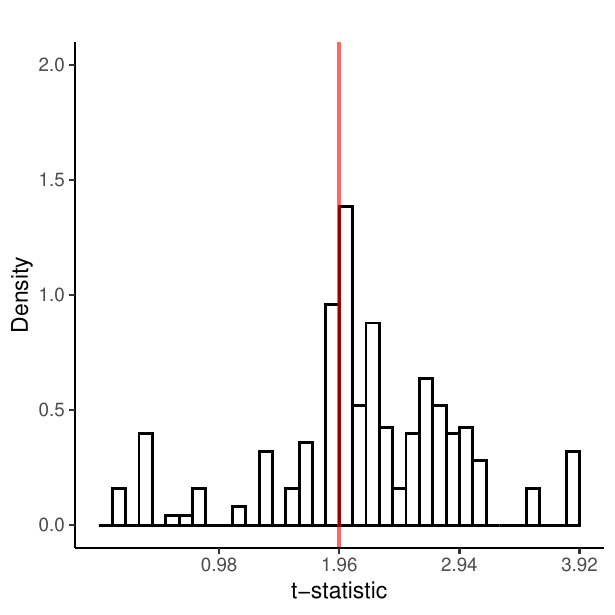}
		\caption{Original approach}
		\label{fig:orig_tstat_noncctrean_all}
	\end{subfigure}%
	\begin{subfigure}{.33\textwidth}
		\centering
		\includegraphics[width=.99\linewidth]{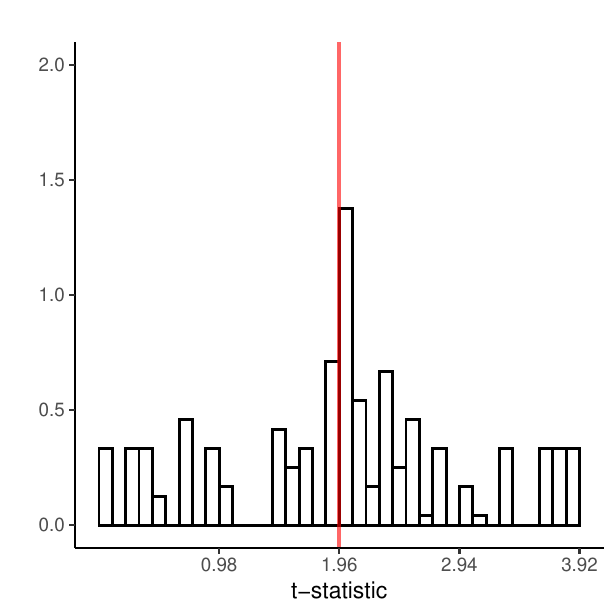}
		\caption{CCT bandwidth}
		\label{fig:classical_tstat_noncctrean_all}
	\end{subfigure}%
	\begin{subfigure}{.33\textwidth}
		\centering
		\includegraphics[width=.99\linewidth]{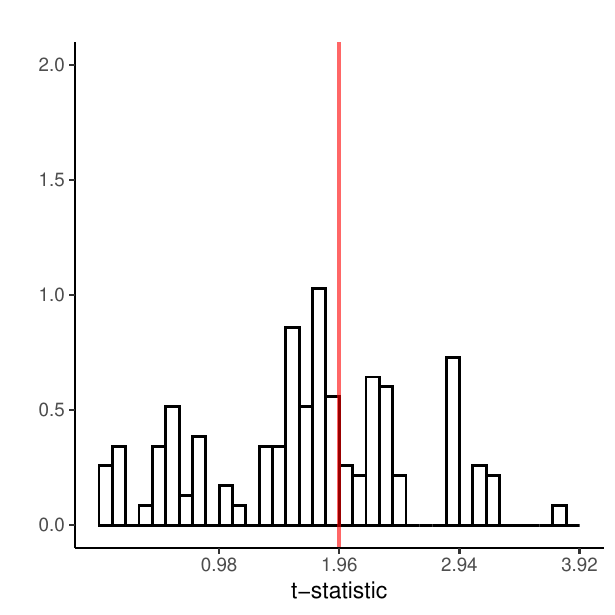}
		\caption{Bias correction and robust SEs}
		\label{fig:cct_tstat_noncctrean_all}
	\end{subfigure}
	\caption{Distributions of $t$-statistics among replicated studies by method of analysis}
	\label{fig:tstat_noncctreanalyzed_all}
\end{figure}

\begin{figure}[H]
	\centering
	\captionsetup[subfigure]{justification=centering}
	\centering
	\begin{subfigure}{.33\textwidth}
		\centering
		\includegraphics[width=.99\linewidth]{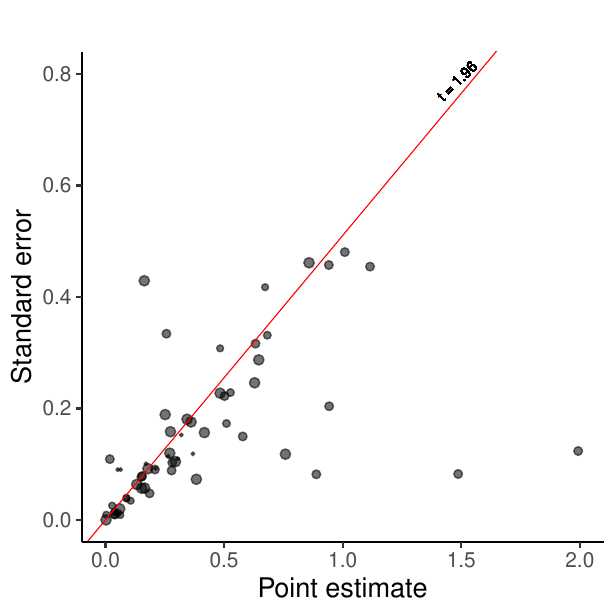}
		\caption{Original approach}
		\label{fig:orig_funnel_noncctrean_all}
	\end{subfigure}%
	\begin{subfigure}{.33\textwidth}
		\centering
		\includegraphics[width=.99\linewidth]{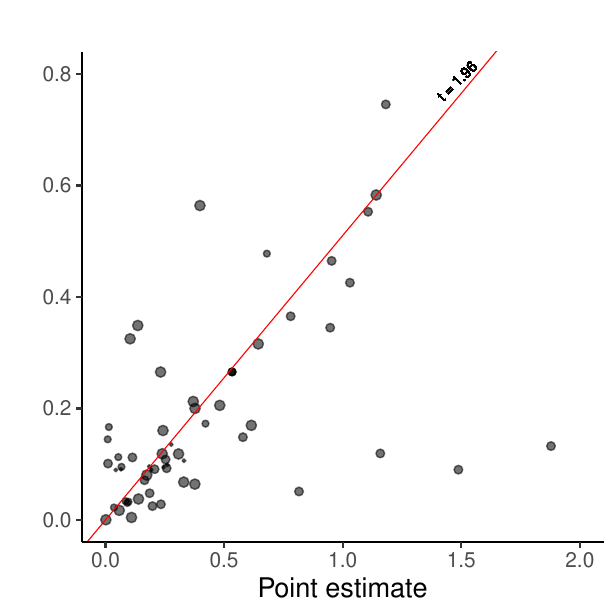}
		\caption{CCT bandwidth}
		\label{fig:classical_funnel_noncctrean_all}
	\end{subfigure}%
	\begin{subfigure}{.33\textwidth}
		\centering
		\includegraphics[width=.99\linewidth]{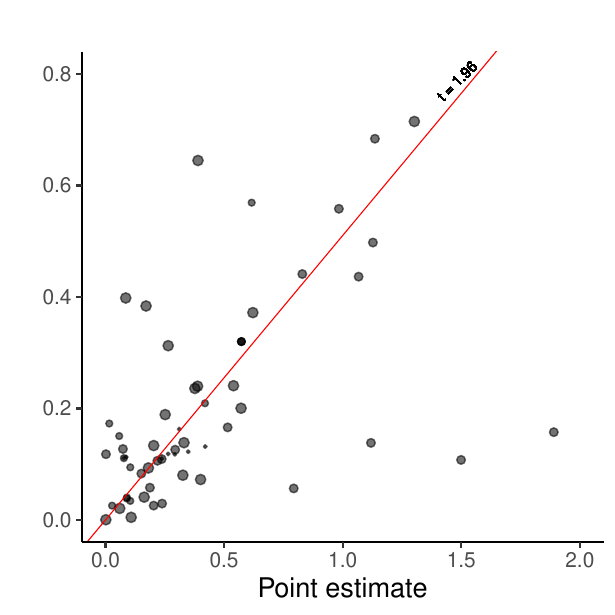}
		\caption{Bias correction and robust SEs}
		\label{fig:cct_funnel_noncctrean_all}
	\end{subfigure}
	\caption{Funnel plots for replicated studies by method of analysis}
	\label{fig:funnel_plots_noncctreanalyzed_all}
\end{figure}

\clearpage

\section{Re-analysis using \texttt{rdhonest} package}\label{sec:rd-honest-supp}

We also reanalyzed studies using the \texttt{rdhonest} package in \texttt{R}. To choose the bandwidth, we used a global quartic regression to estimate a bound on the second derivative for inference under under second order Hölder class, as per the \texttt{NPR\_MROT.fit} function in \texttt{rdhonest}.
The version of the \texttt{rdhonest} package we used cannot account for statistical clustering, and so we were limited to settings that did not originally use clustered standard errors.
We construct pseudo $t$-statistics for the \texttt{rdhonest} package using the length of the confidence interval that include the bias adjustment.

\begin{figure}[H]
	\centering
	\centering
	\includegraphics[width=.45\linewidth]{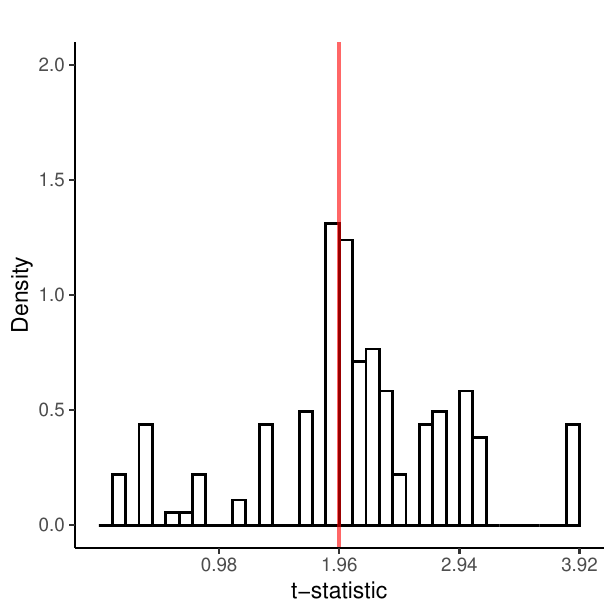}
	\caption{Original $t$-statistics for studies reanalyzed using \texttt{RDHonest}}
	\label{fig:rdhonest-original-tstats}
\end{figure}

\begin{figure}[H]
	\centering
	\centering
	\includegraphics[width=.45\linewidth]{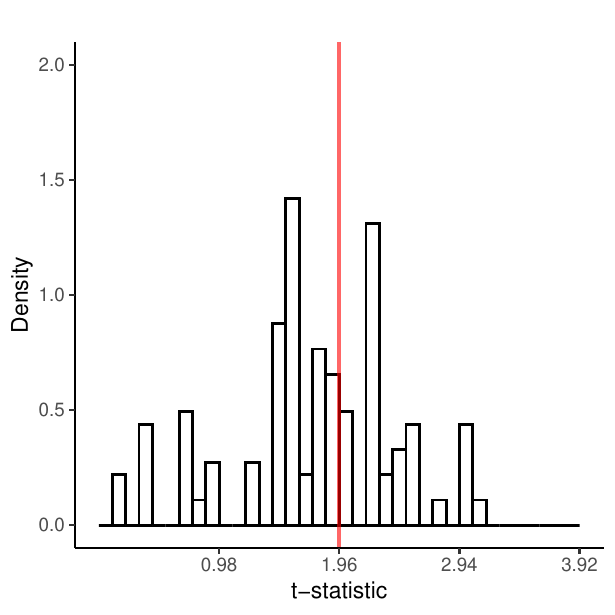}
	\caption{Reanalysis results: $t$-statistics for studies reanalyzed using \texttt{RDHonest}}
	\label{fig:rdhonest-reanalysis-tstats}
\end{figure}

Figure~\ref{fig:rdhonest-original-tstats} presents the original $t$-statistics for the studies we reanalyzed using \texttt{rdhonest}, and Figure \ref{fig:rdhonest-reanalysis-tstats} presents the distribution of $t$-statistics resulting from the \texttt{rdhonest} reanalysis. The reanalyses result in a leftward shift of these estimates.

\begin{figure}[H]
	\centering
	\centering
	\includegraphics[width=.45\linewidth]{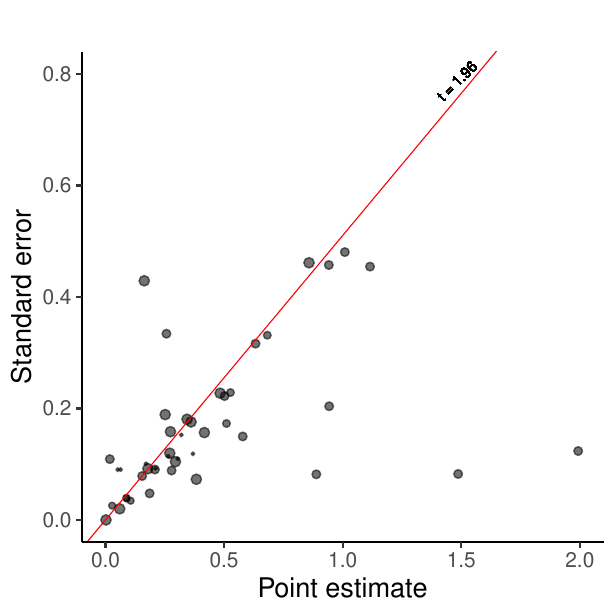}
	\caption{Original funnel plot for studies reanalyzed using \texttt{RDHonest}}
	\label{fig:rdhonest-original-funnel}
\end{figure}

\begin{figure}
	\centering
	\centering
	\includegraphics[width=.45\linewidth]{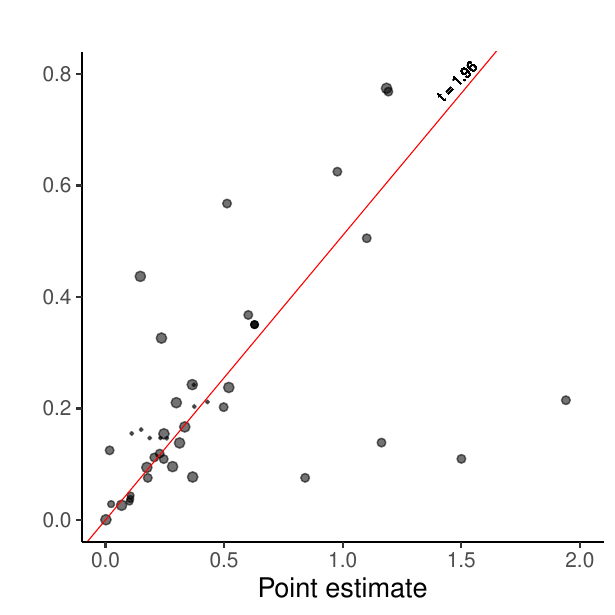}
	\caption{Reanalysis results: funnel plot for studies reanalyzed using \texttt{RDHonest}}
	\label{fig:rdhonest-reanalysis-funnel}
\end{figure}

Figures~\ref{fig:rdhonest-original-funnel}~and~\ref{fig:rdhonest-reanalysis-funnel} are funnel plots similar to those in the main paper but using the \texttt{rdhonest} reanalysis.
Figure~\ref{fig:rdhonest-original-funnel} presents the original results for the subsample of studies included in the \texttt{rdhonest} reanalysis.
Figure~\ref{fig:rdhonest-reanalysis-funnel} presents the results from the \texttt{rdhonest} reanalysis.

Finally, we compare the original, \texttt{rdrobust}, and \texttt{rdhonest} $t$-statistics in Table~\ref{tab:dist-tstats-allapproaches}.
This table is restricted to main estimates, which means that the study by \citet{DAHLGAARD:2018aa} is excluded from rows 2 to 5 because their main estimate is a precision weighted estimate that was not part of our re-analysis.
Both the \texttt{rdrobust} and \texttt{rdhonest} packages result in a left-ward shift of the $t$-statistics.

\begin{figure}[H]
	\centering
	\centering
	\includegraphics[width=.45\linewidth]{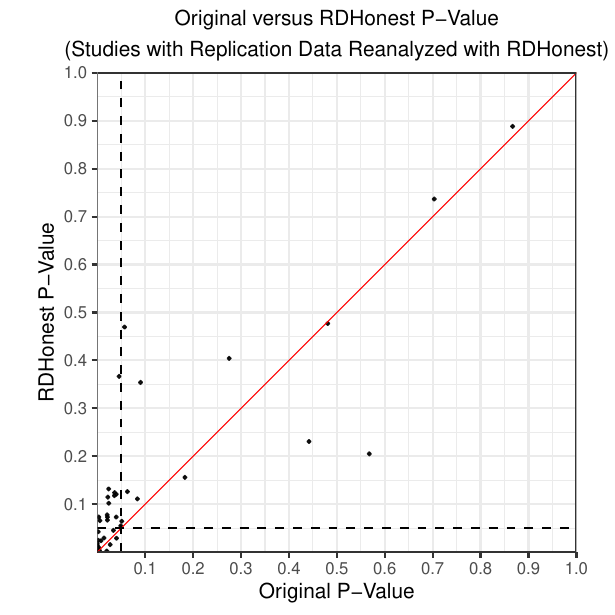}
	\caption{Original $p$-values vs. $p$-values derived using \texttt{RDHonest}}
	\label{fig:rdhonest-pval-comparison}
\end{figure}

\begin{table}[H]
	\footnotesize
	\setlength{\tabcolsep}{10pt}
	\centering
	\caption{Original $t$-statistics versus \texttt{rdrobust} and \texttt{rdhonest}}\label{tab:dist-tstats-allapproaches}
	\resizebox{0.98\textwidth}{!}{%
		\begin{tabular}{lccccccccc}
			\tabinput{tables/tableS6-t-stat-dists.tex}
		\end{tabular}
	}
\end{table}

\end{document}